\documentclass[twocolumn,showpacs,preprintnumbers,amsmath,amssymb]{revtex4}
\usepackage{graphicx}
\usepackage{dcolumn}
\usepackage{bm}
\begin{document}
\preprint{APS}
\title{Self-energy limited ion transport in sub-nanometer channels}
\author{Douwe Jan Bonthuis$^1$}
\author{Jingshan Zhang$^2$}
\author{Breton Hornblower$^{1,3}$}
\author{J\'{e}r\^{o}me Math\'{e}$^1$}
\author{Boris I. Shklovskii$^2$}
\author{Amit Meller$^{1*}$}
 \affiliation{{$^{1}$ Rowland
Institute at Harvard, Harvard University, Cambridge,
Massachusetts 02142}\\
{$^{2}$ William I. Fine Theoretical Physics Institute, University of
Minnesota, Minneapolis, Minnesota}
\\
{$^{3}$ Department of Chemistry and Biochemistry, University of
California at Santa Cruz, Santa Cruz, California 95060}}

\begin{abstract} The current-voltage characteristics of the
$\alpha$-Hemolysin protein pore during the passage of
single-stranded DNA under varying ionic strength, $C$, are studied
experimentally. We observe strong blockage of the current, weak
super-linear growth of the current as a function of voltage, and a
minimum of the current as a function of $C$. These observations
are interpreted as the result of the ion electrostatic self-energy
barrier originating from the large difference in the dielectric
constants of water and the lipid bilayer. The dependence of DNA
capture rate on $C$ also agrees with our model.
\end{abstract}
\pacs{87.14.Gg, 87.15.Tt}

\maketitle

The voltage-driven translocations of
polynucleotides through nanoscale pores has been recently studied
\textit{in vitro} at the single molecule
level~\cite{Kasianowicz1996,Meller2001}. The dynamics of
biopolymer translocation through nanopores is central to many
biological processes such as RNA export and phage infection, and
is the underlying principal behind a number of new methods for
nucleic acids analysis.  Single molecule DNA and RNA translocation
experiments, as well as theoretical models and simulations have
provided some critical information on the dynamics of biopolymer
transport and its dependence on physical parameters such as the
electric field intensity, polymer length, temperature, and
chemical characteristics, such as its
sequence~\cite{Kasianowicz1996,Meller2001,Meller2003}. However,
much less is known about the nature of the \textit{ion current}
during the translocation of the biopolymer through the nanopore.

We use a single $\alpha$-Hemolysin ($\alpha$-HL) pore embedded in
an insulating phospholipid membrane. An ion current of $I \sim 80$
pA is reduced down to $I_B \sim 7$ pA upon the electrophoretic
threading of single stranded DNA molecule into the pore (Fig. 1),
with the ratio $I_B/I =0.09$~\cite{foot5}. Under these conditions
the translocation time of a single base in the polynucleotide is
$\sim$5 $\mu$s~\cite{Meller2001}. To establish a current of $\sim
7$ pA approximately $ 220$ ions must flow through the pore in the
opposite direction during the passage of a nucleotide. As a first
approximation we can therefore assume that the DNA is nearly
static compared with the fast moving ions.

Why is $I_B/I$ so small? The first possible explanation is that
the passage of the ssDNA strongly reduces the cross section of the
channel available for ionic movement. In a bulk solution ssDNA has
the tendency of base stacking leading to formation of a helix with
a diameter of $\sim1\,$nm~\cite{Saenger1984}, which is considerably smaller
than the averaged channel diameter $~1.7$ nm~\cite{Song1996}). Therefore,
such a dramatic current blockage is unlikely to be explained \textit{solely} by the
reduced cross section.

\begin{figure}
\includegraphics[width=6cm,keepaspectratio]{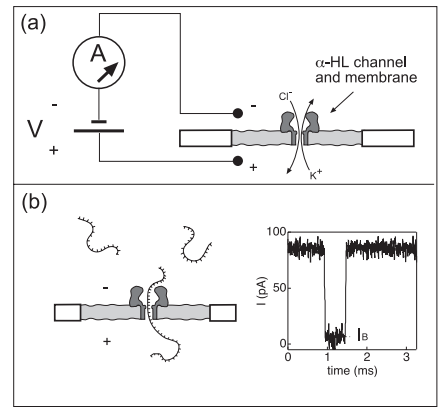}
\caption{\label{fig1} I-V measurements (a) for an ``open" pore,
and (b) when single stranded DNA is electrophoretically threaded
through the pore. DNA abruptly blocks the ion current from $80$ pA
to $\sim 7$ pA ($I_{B}$), as shown in the inset ($C = 1$ M KCl,
V=120 mV, $T=8^{o}$C).}
\end{figure}

In this paper we study how the blocked current $I_B$ depends on
ionic strength $C$ and voltage $V$. We also find that the capture
rate of DNA in the pore strongly depends on $C$.  We explain these
observations as a consequence of an electrostatic self-energy
barrier~\cite{Parsegian} related to the huge difference between
dielectric constant $\kappa_1 = 80$ of water and that of lipids
and DNA $\kappa_2 \sim 2$. This difference results in the
confinement of the electric field lines of an ion during the ssDNA
passage through the channel, leads to a large self-energy of the
ion, and \textit{electrostatically amplifies} the effect of the
channel narrowing.

An electrostatic self-energy barrier was predicted by
Parsegian~\cite{Parsegian}, but its effects have not been directly
observed in the conduction of biological channels. Apparently,
evolution has used several compensating mechanisms to facilitate
ionic transport (screening by salt, ``doping" channels walls by
fixed charges and coating of walls by hanging
dipoles)~\cite{Doyle,Roux,Schuss,Kamenev}. On the other hand,
electrostatic barrier was observed in conduction of synthetic ion
channels~\cite{Lear}, which do not experience evolutionary
pressure. Similarly, $\alpha$-HL is not known to function as a DNA
transporter \textit{in vivo}. Thus, one might expect to observe
effects of the self-energy barrier.

Current-voltage ($I-V$) measurements of a unitary $\alpha$-HL
were performed using dynamic voltage control~\cite{Bates2003,Mathe2005}. Fig.~2
displays the current for the open pore as a function of $C$
measured at 120 mV, 150 mV and 180 mV. In all cases the current
follows a linear dependence on $C$ (thin lines), in the measured
range ($0.25\,$M - $2\,$M). The inset displays the I-V
characteristic curve measured at $C$= 1M KCl.

\begin{figure}
\includegraphics[width=6cm, keepaspectratio]{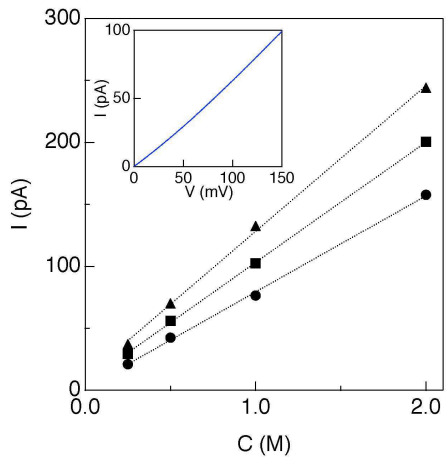}
\caption{\label{fig2} Open pore current characteristics at $T=
8^{o}$C. The main figure displays the dependence of the current on
the bulk KCl concentration measured at $120 mV$, $150 mV$ and $180
mV$ (circles, squares and triangles respectively). Lines are
linear regression fits. The inset is the measured I-V at 1 M KCl.}
\end{figure}

The blocked ion current, $I_{B}$ (Fig.~3), displays several marked
differences as compared to the open pore current. In contrast to
the linear dependence of the open pore current on $C$, the blocked
current is not monotonic. It grows with $C$ roughly linearly for
$C \!\geq\! 1$ M, while at $C \!<\! 1$ M it changes only weakly
and goes  through shallow minimum near $C = 0.5$ M. Furthermore,
the $I-V$ curve of $I_{B}$ measured at 1M KCl (inset of Fig.~3) is
more non-ohmic than that of the open pore current displayed in the
inset of Fig.$\,\,$2.

\begin{figure}
\includegraphics[width=7cm, keepaspectratio]{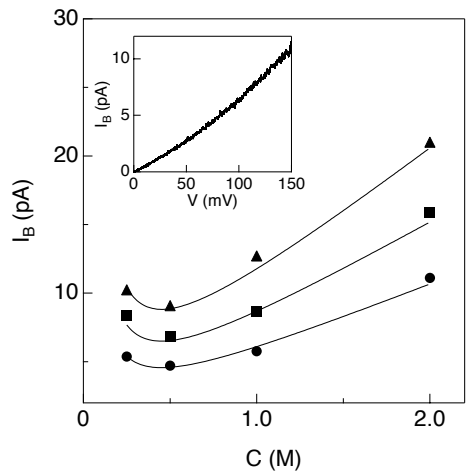}
\caption{\label{fig3} Main figure: The dependence of the blocked ion current,
$I_{B}$ on bulk ion concentration measured at $120 mV$, $150 mV$ and $180
mV$ (circles, squares and triangles respectively).
$I_{B}$ values were determined from the peak of the distributions
of $> 1000$ DNA translocation events for each $C$ and $V$. Solid lines are guides to eyes. The inset displays
the measured I-V curve for the blocked pore, performed using
dynamic voltage control \cite{Bates2003,Mathe2005}.}
\end{figure}

The dependence of the DNA capture rate on the salt concentration
is displayed in Fig.~4. Previous studies showed that the capture
rate scales linearly with the bulk DNA concentration under similar
conditions~\cite{Henrickson2000}. The capture rate dependence on
$C$ is highly non-linear: below $0.5\,$M KCl the capture rate
sharply decreases, and for $C \!>\! 0.5\,$M it levels off. This
trend is consistent for the three different voltages that we
measured. Measurements below 0.25 M were impractical due to the
long delay time between events.

\begin{figure}
\includegraphics[width=6cm, keepaspectratio]{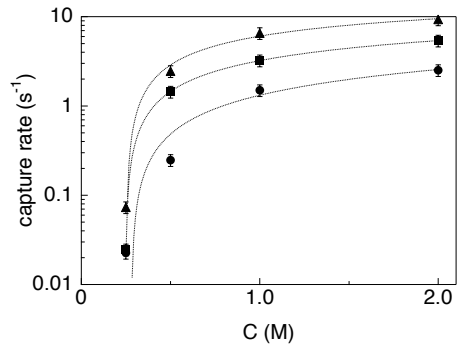}
\caption{The dependence of DNA capture rate (per mole) on KCl
concentration, $C$. Circles, squares and triangles correspond to
$V=120 mV$, $150 mV$ and $180 mV$,respectively} \label{figRate}
\end{figure}

 We use a crude model to
qualitatively interpret the experimental results. Our model is
lacking sophistication and quantitative approach of modern
theories of ion channels~\cite{Schuss,Roux}, but emphasizes
two-dimensional specifics (see below) of DNA translocation
physics, which has not been discussed in the literature.

\begin{figure}[ht]
\begin{center}
\includegraphics[width=6cm, keepaspectratio]{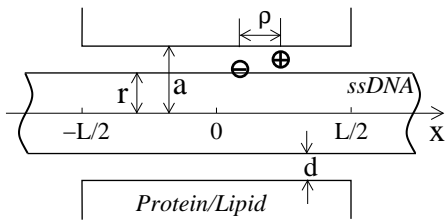}
\end{center}
\caption{The side view of the membrane and the channel with DNA
inserted. A phosphate and a K$^+$ ion bound to it are
shown.}
\label{figdna}
\end{figure}

Let us treat ssDNA and the channel internal wall as coaxial
cylinders (Fig.~\ref{figdna}) with radius $r\!=\!0.5\,$nm and
$a\!=\!0.85\,$nm, respectively. Salt ions are located in the
water-filled space between them, with thickness $d \!=\!
0.35\,$nm. The length of the channel is $L\!=\!5\,$nm.
Each charge on a ssDNA phosphate in the channel is typically
neutralized by a K$^+$ ion, forming a pair. In the confined space
of the ssDNA-occupied pore, K$^+$ ion is strongly bound to a
phosphate moiety on the ssDNA backbone. The strong attraction
arises from confinement of an electrical field between two
cylinders. Within this space opposite charges interact with strong
two-dimensional logarithmic potential $(2e^{2}/\kappa_{1})
\ln(\rho/d)$ (see Fig.~\ref{figfield}), where $\rho\!>\! d$ is the
two-dimensional distance between them. The fact that $d \!\simeq\!
l_B/2$, where $l_B \!=\! e^{2}/\kappa_{1}k_{B}T \!=\!0.7$ nm is
the Bjerrum length and $T$ is the temperature of the experiment,
means that $T \simeq T_c/2$, where $T_c \!=\! e^{2}/(k_B\kappa_1
d)$ is the Kosterliz-Thouless transition temperature~\cite{KT}.
Thus, the thermal motion can not break a pair~\cite{foot4}.

Considering charge transport we may first ignore neutral pairs. An
electric current is produced by an extra ion crossing the channel
in the confined space. Such ion has a higher self-energy than ions
in the bulk solution, or in other words it goes through an
electrostatic self-energy barrier $U(x)$. This blocks the ion
current. To estimate $U(x)$ we write $U(x) \!=\! e\phi(x)/2$,
where $\phi(x)$ is the electrostatic potential created by a charge
$e$ located at $x$. Our numerical calculation in the limit of
infinite ratio $\kappa_{1}/\kappa_{2}$, when all electric lines
stay in the channel and at $a/d\gg 1$ can be well fitted by the
following expression:
\begin{figure}[ht]
\begin{center}
\vspace{0cm}
\includegraphics[width=6cm, keepaspectratio]{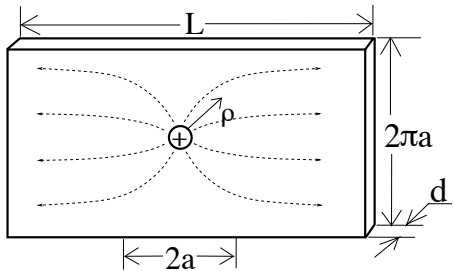}
\end{center}
\caption{An unfolded view of the water-filled space containing an
extra K$^+$ ion. Dashed lines represent the electric field lines
of the charge. At $\rho < a$ this electric field spreads in all
directions and becomes uniform far from the charge.}
\label{figfield}
\end{figure}
\begin{equation}
U(x) \!=\! U_{1}(x)\!+\!U_{2}(x) =\!\frac{e^{2}}{\kappa_1
d}\left[\frac{L}{4a}\left(\!1-\!{4x^2 \over L^2}\!\right) \!+\!
\ln\frac{a}{d}\right]. \label{barrier}
\end{equation}
The origin of the two terms in Eq.~(\ref{barrier}) is illustrated
in Fig.~\ref{figfield} for $x\!=\!0$. At $d \!<\! \rho \!<\! a$
the electric field of the central charge gradually spreads over
all azimuthal angles in the the whole water-filled space decaying
as $E\!=\!e/(\kappa_{1}\rho d)$. This leads to $\phi(x) \!=\!
(2e/\kappa_{1} d)\ln(a/d)$ and the barrier term $U_{2}(x)$. The
$U_2(x)$ is essentially independent of $x$ at the distances larger
than $a$ from the ends and but vanishes at the channel ends, where
most of electric lines are attracted to the bulk solution (this
decay is not reflected by Eq.~(\ref{barrier})). On the other hand,
$U_{1}(0)$ is created by the one-dimensional uniform electric
field at distances $\rho \!>\! a$. For $|x| \!>\! 0$ the electric
field at the closer end is stronger than that of the other end,
therefore $U_1(x)$ decreases parabolically with $|x|$ and vanishes
at the channel ends~\cite{Kamenev}. For $L\!=\! 5\,$nm,
$a\!=\!0.85\,$nm and $d\!=\! 0.35\,$nm Eq.~(\ref{barrier}) gives
$U(0) \!\simeq\! 4.6\,k_{B}T$, or $U_{1}(0)\!=\! 2.9 \,k_BT$ and
$U_{2}(0) \!=\! 1.7 \,k_BT$. The total barrier $U(x)$ of an extra
K$^+$ or Cl$^-$ ion is shown on Fig.~\ref{figband}a by the upper
curves.

\begin{figure}[ht]
\begin{center}
\vspace{-2cm}
\includegraphics[width=6.5cm, keepaspectratio]{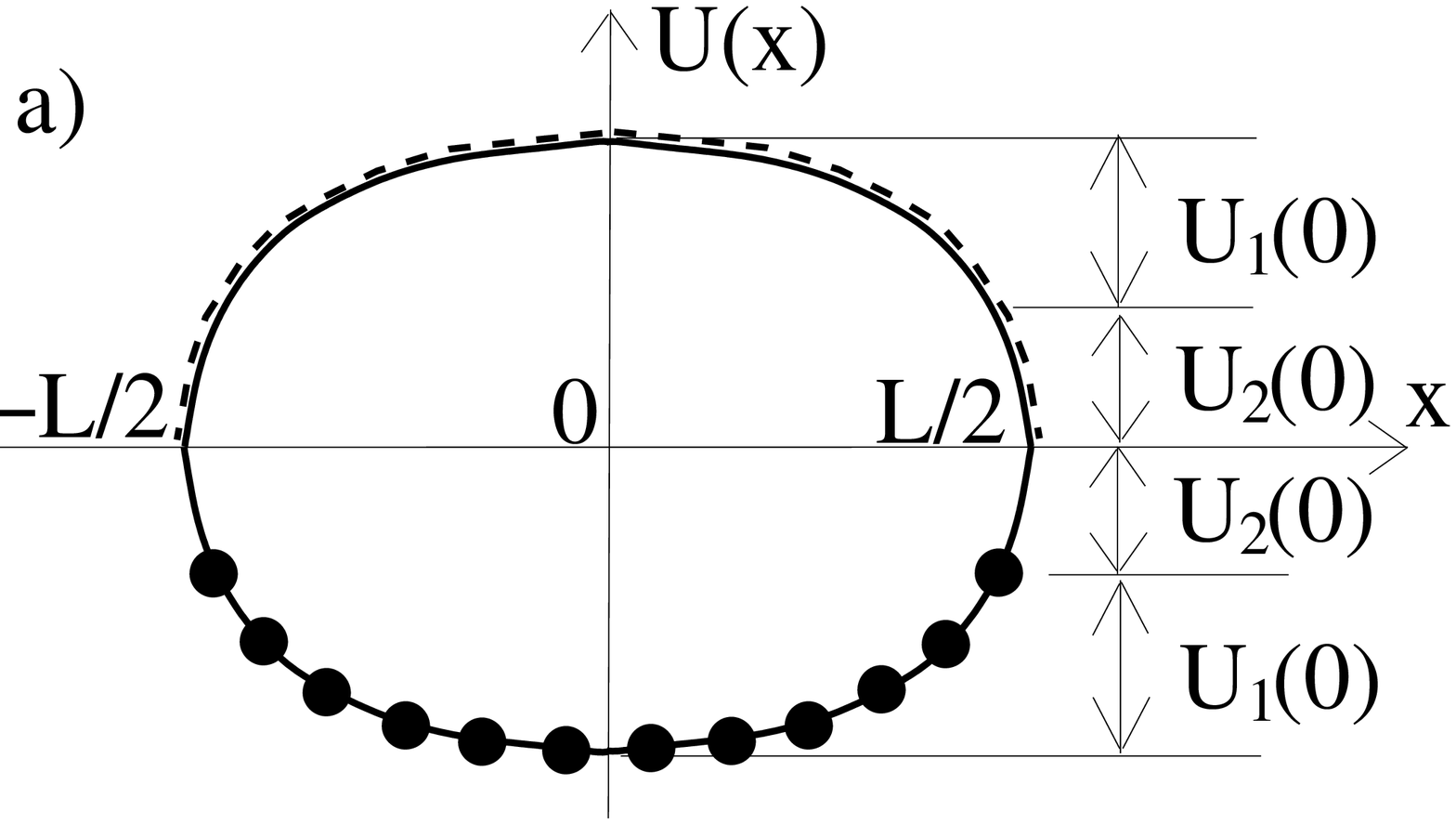}
\end{center}
\begin{center}
\vspace{-4.7cm}
\includegraphics[width=6.5cm, keepaspectratio]{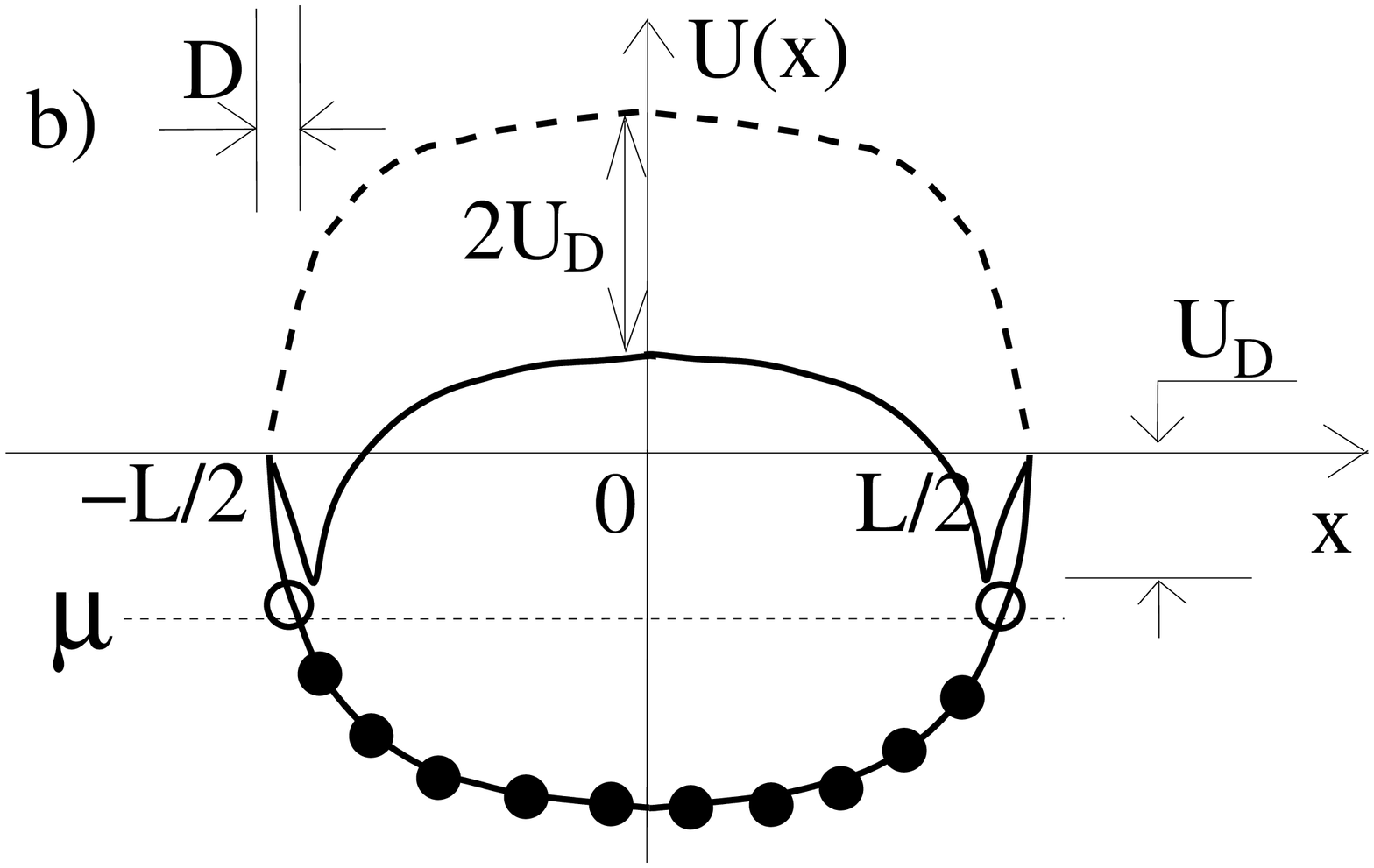}
\vspace{-2cm}
\end{center}
\caption{Energy band diagram for K$^+$ ions (solid lines) and
Cl$^-$ ions (dashed lines). The lower band represents the energy
of the cations bound to DNA phosphates. The empty upper bands show
the self-energy of the extra salt cation (solid line) and anions
(dashed line) entering the channel. a) In the absence of the
Donnan layers ($C \ge C_D$) b) with Donnan layers of the width $D$
creating potential $-U_D$. Vacant phosphates are shown by empty
circles. The chemical potential $\mu$ of K$^+$ ions in the system
is shown by the thin dotted line.} \label{figband}
\end{figure}
Recall that there are K$^+$ ions bound to ssDNA phosphates in the
channel. Each of them can be removed to the bulk creating a
vacancy. The energy penalty for this process is the same as the
penalty for placing an extra ion in the same place. Thus, energies
of bound K$^+$ ions are $- U_{1}(x) -U_2(x)$ and can be shown by
the lower solid curve of Fig.~\ref{figband}a. Vacancies have to
overcome the barrier $U(0)$ to cross the channel.

The large self-energy of extra charges deep in the channel results
in accurate neutralization of DNA by salt cations. Such nearly
perfect neutralization was observed in molecular dynamics
modelling~\cite{Rabin} of the channel.

At small salt concentration $C \!<\! C_D$, some cations close to
the channel ends can escape to the bulk because the cations enjoy
larger entropy in the solution. As a result there are negative
charges in the layer of width $D$ at each end, and positive
screening charge in the adjacent layers of the bulk solution.
These double layers of the width $D$ (see Fig.~\ref{figband}b)
produce the Donnan potential $-U_D$ in the channel, and prevent
remaining cations from leaving the channel. The Donnan potential
moves down energies of both cation bands, while bending these
bands up in the very ends (Fig.~\ref{figband}b). On the other
hand, the energy band of an anions is moved up. This leads to the
exclusion of anions from the channel noticed in Ref.~\cite{Rabin}.

These ideas can be used to interpret the $I_B(C)$ curves in
Fig.~\ref{fig3}. At large enough salt concentration of the bulk
solution $C \!>\! C_D$, where $C_D \!\sim\! 1$ M, ion current
through the channel should be due to extra salt cations and
anions. Therefore the blocked current is proportional to salt
concentration $I_{B} \!\propto\! C$. On the other hand, at $C
\!<\! C_D$ the Donnan potential repels anions and the charge
transport is due to cations only. In the first approximation the
current $I_{B}$ is independent of $C$, because Donnan potential
$U_D$ grows with decreasing $C$, resulting in a smaller barrier
$U(0)$ and for cations to compensate for the decreasing $C$. In
the second approximation, at $C \!<\! C_D$ the Donnan potential
$U_D$ changes slightly slower than the chemical potential
$\mu\!=\!k_BT\ln(C/C_D)$. As a result when $C$ decreases the
barrier for vacancies decreases slightly and the barrier for extra
cations slightly increases. This leads to vacancy dominated
transport~\cite{MacKinnon} and explains the weak increase of
$I_{B}(C)$ as $C$ decreases shown in Fig.~\ref{fig3}. In the limit
of small $C$ beyond the measured range, we would expect the
blocked current to recover $I_B \!\propto\! C$, as the Donnan
potential exceeds the original barrier $U(0)$.

Let us switch to the current-voltage characteristics shown in the
inset of Fig.~3. It is super-linear, because the transport is
limited by electrostatic barriers shown on Fig.~7. On the other
hand, this barrier is relatively flat and therefore the
super-linearity should be weak. This agrees with inset of Fig.~3.

Finally let us concentrate on the sharp dependence of DNA capture rate on $C$ (Fig.~\ref{figRate}). It was
suggested \cite{Ambjšrnsson} that DNA loses conformational entropy
during translocation, and therefore the DNA capture rate acquires
a barrier. However, this mechanism cannot explain the observed $C$
dependence, because the
ssDNA persistence length decreases with $C$, making the
conformation barrier larger. We notice that the screening cloud of
the piece of DNA in the channel is squeezed and corresponding loss
of entropy represents a barrier for the DNA capture. This barrier
decreases with $C$ qualitatively explaining the growing
capture rate with $C$.

To summarize, we studied the dependence of the blocked ion current during DNA translocation on ion concentration. Our
experimental results show non-monotonic behavior, which cannot be accounted for by steric blockade alone. We show that
these results may be expalined by the inclusion of the electrostatic energy of ions inside the pore.

We are grateful to A. Kamenev for useful discussions. We
acknowledge support from National Science Foundation grants no.
NIRT-0403891 and PHY-0417067.

* Electronic mail: meller@rowland.harvard.edu


\begin{thebibliography}{99}


\bibitem{Kasianowicz1996} J. J. Kasianowicz, E. Brandin, D. Branton and D. W. Deamer, Proc. Natl. Acad. Sci.
\textbf{93}, 13770 (1996); M. Akeson, D. Branton, J. J. Kasianowicz, E. Brandin
and D. W. Deamer,  Biophys. J., \textbf{77}, 3227 (1999);  A. Meller, L. Nivon, E. Brandin., J. Golovchenko and D.
Branton, Proc. Natl. Acad. Sci.\textbf{ 97}, 1079 (2000).

\bibitem{Meller2001} A. Meller, L. Nivon, D. Branton, Phys. Rev. Lett. \textbf{86}, 3435 (2001).

\bibitem{Meller2003} A. Meller, J. Phys. Cond. Matt. {\bf 15}, R581 (2003).

\bibitem{foot5} This value corresponds to ss-DNA entering the pore with its $3'$ end, which is
the most probable direction \cite{Mathe2005}.


\bibitem{Saenger1984} W. Saenger, Principles of Nucleic Acid
Structure, Springer Verlag New York, (1984)

\bibitem{Song1996} L. Song,  M.R. Hobaugh, C. Shustak, S. Cheley,
H. Bayley, \and J.E. Gouax  , Science, \textbf{274}, 1859 (1996).

\bibitem{Parsegian} A. Parsegian, Nature {\bf 221}, 844 (1969);
P. C. Jordan, Biophys J. {\bf 39}, 157 (1982).

\bibitem{Doyle} D. A. Doyle et al., Science. {\bf 280}, 69 (1998); R.
MacKinnon, Nobel Lecture. Angew Chem Int Ed Engl {\bf 43} 4265
(2004).

\bibitem{Roux} S. Kuyucak, O. S. Andersen, and S-H. Chung,
Rep. Prog. Phys. 64, 1427 (2001); B. Nadler,  U. Hollerbach, R. S.
Eisenberg, Phys. Rev. E {\bf 68}, 021905 (2003); B. Roux, T.
Allen, S. Berneche, W. Im, Quart. Rev. Biophys. {\bf 37}, 15
(2004).

\bibitem{Schuss} G. R. Dieckmann, J. D. Lear, Q. Zhong, M. L. Klein,
W. F. DeGrado, K. A. Sharp, Biophys. J. {\bf 76} 618 (1999); A. B.
Mamonov, R. D. Coaalson, A. Nitzan, M. Kurnikova, Biophys. J. {\bf
84} 3646 (2003); Z. Schuss, B. Nadler, R. S. Eisenberg, Phys. Rev.
E, {\bf 64} 036116 (2003); A. Burykin, A. Warshel, Biophys. J.
{\bf 85} 3696 (2003); P. Graf, M. Kurnikova, R. D. Coalson, A.
Nitzan, J. Phys. Chem. {\bf 108} 2006 (2004)


\bibitem{Kamenev} A.~Kamenev, J.~Zhang, A.~I.~Larkin,
B.~I.~Shklovskii, Physica A 359, 129 (2006); J.~Zhang, A.~Kamenev,
B.~I.~Shklovskii, Phys. Rev. Lett. {\bf 95}, 148101 (2005).

\bibitem{Lear}  J. D. Lear, J. P. Schneider, P. K. Kienker,
W. F. DeGrado, J. Am. Chem. Soc. {\bf 119}, 3212 (1997).

\bibitem{Mathe2005} J. Math\'{e}, A. Aksimentiev, D. Nelson, K. Schulten and A.
Meller, Proc. Natl. Acad. Sci., \textbf{102}, 12377 (2005).

\bibitem{Bates2003} M. Bates, M. Burns, A. Meller, Biophys. J.,
\textbf{84}, 2366 (2003).


\bibitem{Henrickson2000} S. E. Henrickson, M. Misakian, B. Robertson
\and J. J. Kasianowicz, Phys. Rev. Lett. \textbf{85} 3057 (2000);
A. Meller \and D. Branton Electrophoresis \textbf{23}, 2583
(2002).

\bibitem{KT} J.M. Kosterlitz, D.J. Thouless, J. of Phys. C: Sol.
St. Phys., {\bf 6} 1181 (1973).

\bibitem{foot4} One may wonder what happens if the whole DNA
cylinder is shifted to one side. It easy to show that because of
spiral location of charges this will leave cations localized.

\bibitem{Rabin} Y. Rabin, M. Tanaka, Phys. Rev. Lett.
{\bf 94}, 148103 (2005).

\bibitem{MacKinnon} M. F. Schumaker, R. MacKinnon,
Biophys. J. {\bf 58}, 975 (1990).

\bibitem{Ambjšrnsson} T. Ambj\"{o}rnsson, S. P. Apell, Z. Konkoli,
E. A. Di Marzio, and J. J. Kasianowicz, J. Chem. Phys.
\textbf{117}, 4063 (2002); W. Sung and P. J. Park, Phys. Rev.
Lett. \textbf{77}, 783 (1996); M. Muthukumar, J. Chem. Phys.
\textbf{111} 10371, (1999).

\end{thebibliography}
\end{document}